\documentclass[10pt,a4paper,twocolumn,aps,prl,nobibnotes,floatfix,superscriptaddress,showpacs]{revtex4-1}

\usepackage[utf8]{inputenc}
\usepackage[english]{babel}

\usepackage{amsfonts}
\usepackage{amsmath}
\usepackage{amssymb}
\usepackage{bbold}
\usepackage[dvipsnames,svgnames,table]{xcolor}
\usepackage{graphicx}
\usepackage{fancyhdr}
\usepackage{float}
\usepackage{braket}
\usepackage{float}
\usepackage[caption=false]{subfig}
\usepackage{tikz}  % drawings
\usepackage{chngcntr}

\usepackage{hyperref}
\hypersetup{
    pdfstartview={FitH},% fits the width of the page to the window
    colorlinks=true,    % false: boxed links; true: colored links
    linkcolor=NavyBlue, % color of internal links
    citecolor=Maroon,   % color of links to bibliography
    filecolor=NavyBlue, % color of file links
    urlcolor=NavyBlue   % color of external links
}

\makeatletter
\def\footnoterule{\kern -10pt
    \hrule \@width 100pt \kern 10pt} % the \hrule is .4pt high
\makeatother

\begin{document}

\title{Tuning domain wall oscillation frequency in bent nanowires through a mechanical analogy}
% Force line breaks with \\

\author{G. H. R. Bittencourt}
\author{V. L. Carvalho-Santos}
\affiliation{Departamento de F\'isica, Universidade Federal de Vi\c cosa, Av. PH Rolfs s/n, 36570-900, Vi\c cosa, Brazil}

\author{D. Altbir}
\affiliation{Universidad Diego Portales, Ej\'ercito 441, Santiago}

\author{O. Chubykalo‐Fesenko}
\affiliation{Instituto de Ciencia de Materiales de Madrid, CSIC, Cantoblanco, 28049 Madrid, Spain}

\author{R. Moreno}
\affiliation{Instituto de Física Enrique Gaviola, IFEG (UNC-CONICET), Medina Allende s/n, Ciudad Universitaria, 5000 Córdoba, Argentina}

\email{vagson.santos@ufv.br}

\begin{abstract}

 In this work, we present a theoretical model for domain wall (DW) oscillations in a curved  magnetic nanowire with a constant curvature under the action of a uniaxial magnetic field. Our results show that the DW dynamics can be described as that of the mechanical pendulum, and both the NW curvature and the external magnetic field influence its oscillatory frequency. A comparison between our theoretical approach and experimental data in the literature shows an excellent agreement. The results presented here can be used to design devices demanding the proper control of the DW oscillatory motion in NWs.
\end{abstract}

\maketitle

Magnetic nanowires (NWs) are elongated nanostructures with potential applications in several   research fields \cite{hilosMateriales,hilosAgua,hiloshyperthermia,hilospalaeomagnetismo}. Their magnetization reversal is a hysteretic process that occurs through the nucleation of magnetic domain walls (DWs) that propagate along the NW length\cite{IvanovReversalModes}. These DWs can exhibit different features depending on the geometrical and magnetic parameters of the NW \cite{MORENO2022168495}. Therefore, engineering the NW geometry is a promising way to the proper control of the DW dynamical properties, crucial for a broad range of different applications \cite{Parkin190,Sharma2015,doi:10.1063/1.4881061,Allwood2003,Alvaro,Voto,Berger,Hung,Yamanouchi2004,PhysRevLett.93.127204}.

Current  fabrication techniques at the nanoscale allow the preparation of NWs with well-controlled shapes and sizes \cite{ExpAlbrecht2005,ExpMinguez-Bacho_2014,ExpRollingTubes,ExpSphericalCaps,ExpTejo2020,Ref1,Ref2,Ref3,Ref4,Ref5}.  Particularly, advances in 3D-printing techniques \cite{Pachecho2020,helices} have enhanced the fabrication of multifaceted \cite{SANZHERNANDEZ202085} and curved \cite{Sanz-Hernández2020,ExpFernández-Pacheco2013,locomotion} magnetic NWs. Such experimental techniques would allow corroborating theoretical results that predict that DWs displacement in multifaceted and bent NWs under the action of a constant external stimulus  exhibit an oscillatory behavior \cite{Altbir2020,PhysRevB.96.184401,secondWalker,Yershov-Helice}. Indeed, transverse DWs propagating in straight NWs with rectangular cross-sections simultaneously rotate and oscillate around and along the NW \cite{OsciStripes} when a magnetic field is applied parallel to its axis. This phenomenon occurs above a threshold for the external stimuli known as Walker breakdown \cite{Walker,Mougin_2007},  originating on changes in the magnetic energy due to the existence of hard axial \cite{Walker} or shape \cite{Mougin_2007} anisotropies, that define a preferential DW phase (direction along which the DW magnetic moments point). On the other hand, the DW phase is arbitrary in straight NWs with cylindrical cross-section and zero magnetocrystalline anisotropy, and no oscillations in the DW position appear \cite{Hertel}. Nevertheless, the DW oscillation can be recovered by bending the cylindrical NW \cite{PhysRevB.96.184401,Yershov-Helice,secondWalker,WalkerhilosCurvos}. The amplitude and frequency of these oscillations can be controlled in terms of several parameters, such as the cross-section area, the magnetic materials of the NW, and its curvature.

To better understand DW dynamics, mechanical analogies are often used  considering DWs as a particle-like structure \cite{relativity1,Cherenkov}. Under this frame, the DW oscillatory motion in bent NWs with constant curvature can be described starting with the analysis of the angular momentum of the DW in analogy with a body under the action of a central force. This mechanical analogy associates some constants of motion to the DW displacement \cite{AreaLaw}. 
This assertion does not hold for bent NWs with a curvature gradient. In these nanostructures, pinning effects at the maximum curvature point \cite{Volkov-PRL,Lewis-APL,CP1,CP2} produces an oscillatory decay of the DW position \cite{Yershov-Pinning,hilosElipticos}. This behavior can be described by comparing the system with a spring-mass oscillator with the restoring field and DW oscillation frequency depending on the curvature  gradient \cite{hilosElipticos}. 

Mechanical analogies for the DW dynamics have also been used to phenomenologically explain results obtained by Saitoh \textit{et al.}, who experimentally determined the DW mass by measuring the resonant frequency of a DW in a bent NW as a function of the applied field magnitude. In that case, the authors compared the DW dynamics to the oscillation of a single pendulum, assuming that the magnetic field applied to the wire plays the role of a gravitational field \cite{Saitoh2004}. The proposed mechanical analogy allowed the authors to obtain estimations of the energy and oscillation frequency of the DW motion. 

\begin{figure}[!tbp]
    \centering
    \includegraphics[width=8cm ,angle=0]
    {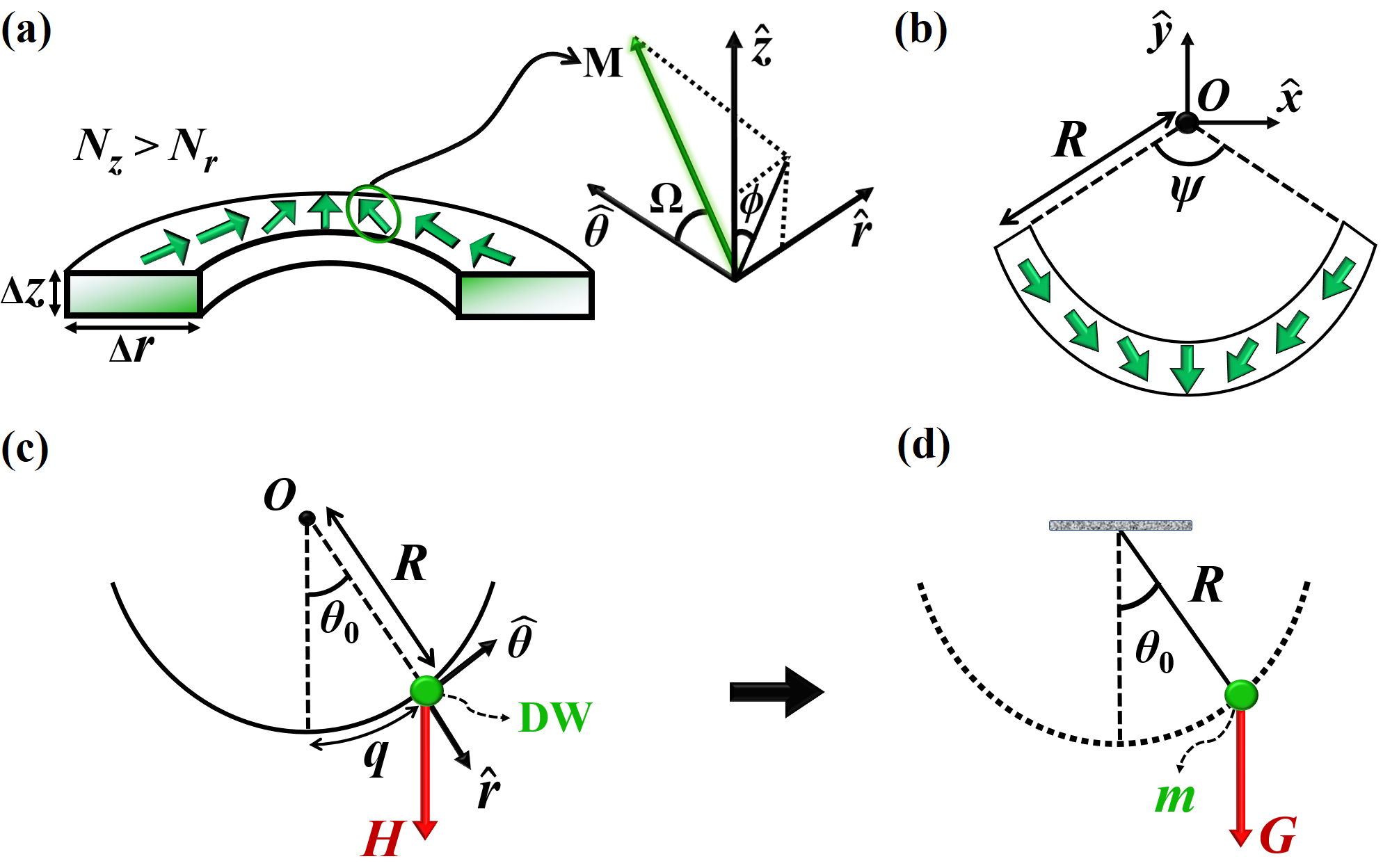}
    \caption{\textbf{(a)} Schematic representation of a DW and the adopted coordinates systems. \textbf{(b)} Illustration of the magnetization profile and the Cartesian axes. \textbf{(c)} and \textbf{(d)} depict the analogy between the DW lying on a bent NW under a homogeneous field $\boldsymbol{H}$ and a simple pendulum under the action of the gravitational force $\boldsymbol{G}$.}
    \label{fig1}
    \end{figure}

In this work, we go deeper into this analysis by using a collective variables approach to study the DW dynamics in bent NWs under the action of a uniaxial magnetic field, as shown in Fig. \ref{fig1}. We demonstrate that our rigorous approach can be reduced to equations equivalent to the phenomenological one of Saitoh \textit{et al.} \cite{Saitoh2004}. In addition, we also show that the oscillatory regime can be over or underdamped, depending on the magnetic field strength and NW curvature. Finally, our results allow us to define  the DW mass, giving values that agree well with the experimentally measured by Saitoh \textit{et al.} \cite{Saitoh2004}. The detailed description of the DW dynamics presented in this work is interesting for applications in nano-oscillators \cite{Sato-PRL,Osc-JAP,Osc-Nat,Osc-Rev} and DW-based neuromorphic \cite{Neuro1,Neuro2,Neuro3}.

Our system consists of an NW with a rectangular cross-section defined by $\Delta r$ (width) and $\Delta z$ (thickness). Additionally, the NW has a constant curvature $\kappa = 1/R$, where $R$ is the radius of a circle with origin at $O$, as depicted in Fig. \ref{fig1}. The NW is subjected to a homogeneous magnetic field $\boldsymbol{H} = -H \hat{y}$, which, in its mechanical analogous, is a gravitational force, $\boldsymbol{G}$. 
The magnetization profile is parameterized in a curvilinear basis as $\boldsymbol{m}=\sin\Omega\sin\Phi\,\hat{r}+\cos\Omega\,\hat{\theta}+\sin\Omega\cos\Phi\,\hat{z}$,
 being $\Omega \equiv \Omega(\theta)=2\arctan\left\{\exp[p(s-q)/\lambda]\right\}$, where $\lambda$ is the DW width parameter, $s = R\theta$ defines an arbitrary coordinate on the NW, and $q = R\theta_0$ represents the position of the DW center. $p = +1$ ($p = -1$) corresponds to a head-to-head (tail-to-tail) DW. For now on, and for simplicity, we state $p = +1$. However, all results presented here are promptly recovered by replacing $p\rightarrow-p$ and $\boldsymbol{H}\rightarrow-\boldsymbol{H}$. Furthermore, as shown in \cite{Yershov-Pinning}, the rigid DW model allows the representation of the direction at which the DW center points (DW phase) as $\Phi=\phi$, which is independent of $\theta$.

To understand the DW steady states, we determine the total energy per cross-section area, $E = \int \mathcal{E} R d\theta$, where the energy density $\mathcal{E} = \mathcal{E}_x + \mathcal{E}_d + \mathcal{E}_Z$ contains the exchange, dipolar, and Zeeman contributions. The exchange energy density can be written as $\mathcal{E}_x = A \left(\kappa \, \partial \boldsymbol{m}/ \partial \theta \right)^2$, where $A$ is the exchange stiffness and $\boldsymbol{m}=\boldsymbol{M}/M_s$ is the normalized magnetization, with $M_s$ being the saturation magnetization.  From now on, results presented in this work consider a Permalloy NW ($A = 1.3 \times 10^{-6}$ erg/cm and $M_s = 795$ emu/cm$^{3}$) with cross-section dimensions $\Delta r = 20$ nm and $\Delta z = 10$ nm, ensuring the creation of a transverse domain wall during magnetization reversal. The shape anisotropy approximation  allows writing the dipolar energy as $\mathcal{E}_d = 2\pi M_s^2 \sin^2\Omega\left(N_r\sin^2\phi + N_z\cos^2\phi \right)$, where $N_r$ and $N_z$ are the demagnetizing factors associated to the $\hat{r}$ and $\hat{z}$ directions, respectively.  Here, $N_r$ and $N_z$ are determined from the reduction of the magnetostatic energy of a straight and uniformly magnetized stripe with a rectangular cross-section to the effective shape anisotropy \cite{Aharoni}. This approximation can also be used for thin, narrow, and curved stripes hosting inhomogeneous magnetization textures \cite{Gaididei-JPA}, including domain
walls \cite{Mougin_2007,secondWalker,Yershov-Auto}. Finally, the Zeeman contribution to the magnetic energy is $\mathcal{E}_Z=-\boldsymbol{M}\cdot\boldsymbol{H}$, with $\boldsymbol{H} = H(\cos\theta \hat{r}-\sin\theta \hat{\theta})$. In this model, the exchange and dipolar energy densities can be integrated along the NW length $L$, giving

\begin{equation}
    E_x =A\left[\frac{2}{\lambda}-2\pi\kappa\sin\phi-\kappa^2(\lambda+\lambda\cos2\phi-L)\right] \, ,
    \label{Ex}
\end{equation}

\noindent and 

\begin{equation}
    E_d =4\pi\lambda M_s^2(N_r \sin^2\phi+N_z\cos^2\phi) \, .
    \label{Ed}
\end{equation}

The integration of $\mathcal{E}_Z$ yields an expression given by a series of hypergeometric functions, similar to those obtained in Ref\cite{PhysRevB.96.184401}. Nevertheless, and assuming that the DW is far from the NW borders ($q \ll L$),  the DW width ($\Delta_{_{W}}=\pi\lambda$) is much smaller than the NW length ($L \gg \Delta_{_{W}}$), and discarding all energy constants (terms that do not depend on $q$ and $\phi$), the expression for the Zeeman energy can be expanded up to terms quadratic in $q$, giving

\begin{equation}
    E_Z = M_s H\left[\lambda (4\kappa^2\lambda^2 - \pi)\sin\phi + \left(1+\frac{\pi}{2}\kappa \lambda \sin\phi\right)\kappa q^2\right] \, .
    \label{Ez}
\end{equation}

\begin{figure}[!tbp]
    \centering
    \includegraphics[width=8.5cm ,angle=0]
    {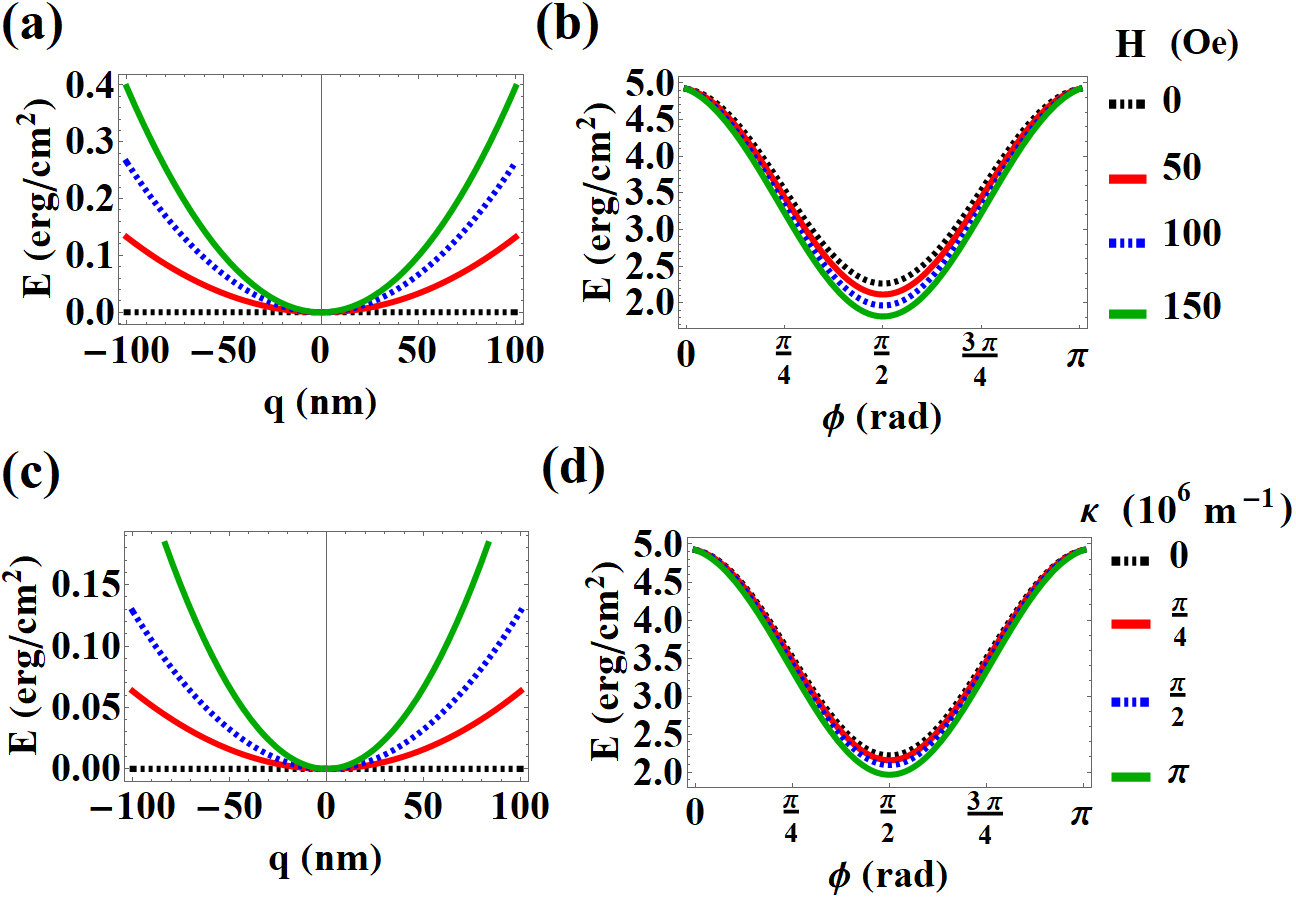}
    \caption{Domain wall total energy $E$ as a function of \textbf{(a)} the DW position for $\phi = \pi/2$,  \textbf{(b)} the phase for fix curvature  $\kappa = \pi \times 10^6$ m$^{-1}$ and $q = 0$, and different $H$ values. \textbf{(c)} and \textbf{(d)} show the DW energy as a function of $q$ and $\phi$, respectively for $H = 100$ Oe and different values of $\kappa$. }
    \label{energy}
    \end{figure}
    
The analysis of the contributions to the magnetic energy reveals that $E_x$ and $E_d$ do not depend on $q$. On the other hand, the last term in Eq. (\ref{Ez}) evidences that the interplay between the magnetic field and the NW curvature generates a dependence of the Zeeman energy on the DW position.  Thus, Zeeman energy creates a harmonic potential for the DW center with an energy minimum  at $q = 0$. Additionally, because we are dealing with a head-to-head DW, the minimum energy occurs when $\phi = \pi/2$ \cite{PhysRevB.96.184401,WalkerhilosCurvos,secondWalker}. Fig. \ref{energy} illustrates the total magnetic energy as a function of the DW position and phase. Figs. \ref{energy}-a and b present the results for an NW with curvature $\kappa=\pi\times10^{6}$ m$^{-1}$ under the action of different magnetic fields, while Figs. \ref{energy}-c and d show the results for NWs with different curvatures under the action of a magnetic field $H=100$ Oe. One can notice that the increase of both $\kappa$ and $H$ yield narrower potential wells  for the DW dynamics. Therefore, the results show that the DW behaves as a particle performing a harmonic oscillation \cite{secondWalker,hilosElipticos}. In this context, we can determine the $\chi$-component of the generalized restoring-like forces per cross-section area, $\mathcal{F}_{\chi}=-\partial_\chi E$, that bring the DW to the equilibrium position. By deriving $E$ with respect to the coordinates $q$ and $\phi$ and considering small displacements, we obtain

\begin{eqnarray}
%    \centering
%    \begin{array}{cl}
   \mathcal{F}_{q} \approx -\mathcal{K}_1 q  
,\hspace{0.3cm}{\text{and}}\hspace{0.3cm}
    \mathcal{F}_{\phi} \approx -\mathcal{K}_2\left(\phi - \dfrac{\pi}{2}\right) \, .
    \label{Forces}
    \end{eqnarray}

\noindent Here, $\mathcal{K}_1$ and $\mathcal{K}_2$ represent the respective elastic-like constants related to the DW translational and rotational motions and are evaluated as

\begin{eqnarray}
    \centering
    \begin{array}{cl}
   \mathcal{K}_1 = M_s H \kappa(2+\pi\kappa\lambda)  \, ,\\
    \\
    \mathcal{K}_2 = 2A\kappa(\pi-2\kappa\lambda) + M_s H \lambda(\pi-4\kappa^2\lambda^2) + 8\pi\lambda M_s^2 \mathcal{N} \, ,
    \end{array}
    \label{elastic-constant}
\end{eqnarray}

\noindent where $\mathcal{N} = N_z - N_r$.

On the other hand, in a more strict approach, the DW dynamics can be determined in terms of the Lagrangian $\mathcal{L} = \mathcal{S}\int \left(\mathcal{E} + \frac{M_s}{\gamma}\dot{\Phi}\cos\Omega\right) R\,d\theta$ and the Rayleigh dissipation function $\mathcal{D} = \mathcal{S}\int \left[\frac{\alpha M_s}{2\gamma}\left(\dot{\Omega}^2 + \dot{\Phi}^2\sin^2\Omega\right)\right] R\,d\theta$, where $\mathcal{S}$ is the NW cross-section area. Under this framework, the dynamical equations are obtained by \cite{Thiavile}

\begin{equation}\label{euler-lagrange}
    \frac{\delta \mathcal{L}}{\delta \chi} - \frac{d}{dt}\left( \frac{\delta \mathcal{L}}{\delta \dot{\chi}}\right) = -\frac{\delta \mathcal{D}}{\delta\dot{\chi}} \, \,  ,  \chi \, \in \{\Omega, \Phi\} \, ,
\end{equation}

\noindent where $\delta/\delta \chi$ and $\delta/\delta \dot{\chi}$ are the variational derivatives. We highlight that Eq. (\ref{euler-lagrange}) is equivalent and can be obtained from the Landau-Lifshitz-Gilbert (LLG) equations. The main dynamical properties of a DW can be obtained from Eq. (\ref{euler-lagrange}) by using a collective variables approach \cite{Slonczewski,Thiavile,Kravchuk}, where the DW profile is assumed to propagate along the NW obeying a traveling wave ansatz described by $\Omega(s, t) = 2\arctan\left\{\exp[(s-q(t))/\lambda(t)]\right\}$ and $\Phi = \phi(t)$. Under these assumptions, the integrals in $\mathcal{L}$ and $\mathcal{D}$ can be solved to obtain the effective functions \cite{Yershov-Helice}

\begin{equation}
    \centering
    \begin{array}{cl}
     \mathcal{L}_{\text{ef}} = \mathcal{S}\left(E+\dfrac{2M_s}{\gamma}q\dot{\phi}\right) \, ,\\
    \\
    \mathcal{F}_{\text{ef}} = \mathcal{S}\dfrac{\alpha M_s}{\gamma}\left(\dfrac{\dot{q}^2}{\lambda}+\lambda\dot{\phi}^2 + \dfrac{\pi^2\dot{\lambda}^2}{12\lambda}\right) \, \, . 
    \end{array}
    \label{LagrangianEff}
\end{equation}

The DW dynamics can be obtained by inserting \eqref{LagrangianEff} in \eqref{euler-lagrange} and solving it for a new set of generalized coordinates $\{q(t), \phi(t), \lambda(t) \}$. Although this theoretical model was first developed for straight systems, it is also valid to determine the DW dynamics in bent NWs \cite{Yershov-Pinning,Yershov-Helice} in the limit $\kappa \lambda \lesssim 1$. Therefore, the range of curvatures here considered allows us to adopt this model to determine the DW dynamics in the considered system. In this context, and following this procedure, we have that

\begin{equation}
    \centering
    \begin{array}{cl}
     \dot{q} = -\dfrac{\gamma}{2M_s} \mathcal{F}_\phi + \alpha \dot{\phi}\lambda \, ,\\
    \\
    \dot{\phi} = \dfrac{\gamma}{2M_s} \mathcal{F}_q - \dfrac{\alpha}{\lambda} \dot{q} \,, \\
   \\
   \dot{\lambda} = -\dfrac{6\gamma\lambda}{\alpha\pi^2M_s}\dfrac{\partial E}{\partial \lambda} \, 
    \end{array}
    \label{eq.motion_q_phi_lambda}
\end{equation}

\noindent describe respectively the DW position, phase, and width as a function of time. The numerical analysis of the above set of equations reveals that after a time interval of $\sim 10^{-11}$ s, $\lambda(t)$ relaxes towards the value 

\begin{equation}
    \lambda = \sqrt{\frac{A}{2\pi M_s^2(N_r \sin^2\phi+N_z^*\cos^2\phi)-\eta_H \sin\phi}}
\label{delta}
\end{equation} 

\noindent 
where $\eta_H = \frac{\pi}{2} H M_s\left(1-\frac{1}{2}\kappa^2q^2 \right)$, $N_z^* = N_z - \ell^2\kappa^2$, $\ell = \sqrt{A/2\pi M_s^2}$ is the exchange length.  Therefore, the DW width can be considered as a position-dependent slave variable ($\lambda = \lambda[q(t), \phi(t)])$. In the system under analysis, the DW oscillates with a small amplitude around the position $q=0$. In addition, the magnetic field, curvature-induced phase selection \cite{Yershov-Pinning}, and the shape anisotropy force the DW to point outward the bent so that $\phi \approx \pi/2$. In this case, we have that $\lambda \approx \sqrt{A/\left(2\pi N_r M_s^2 -  \pi M_s H/2\right)}$ becomes time-independent and can be considered as a constant during all motion. Indeed, although $\lambda$ depends on $H$, its variation for a NW with $\Delta r = 20$ nm and $\Delta z = 10$ nm is in the range of $\lambda \in [11.5, 12.5]$ nm when $H \in [0, 150]$ Oe. Therefore, from now on, we consider that $\lambda$ is constant during the propagation and is given by the averaged value, $\lambda = 12$ nm.  Under these constraints, the substitution of Eq. (\ref{Forces}) in \eqref{eq.motion_q_phi_lambda} yields

 \begin{eqnarray}
    \centering
%    \begin{array}{cl}
     \Ddot{q} = -\dfrac{1}{1+\alpha^2} \left(\omega_0^2 q + 2\beta \dot{q} \right) \, 
         \label{oscillator-eq-q}
\end{eqnarray}
\noindent and

\begin{eqnarray}
   \Ddot{\phi} = -\dfrac{1}{1+\alpha^2} \left[\omega_0^2 \left(\phi-\dfrac{\pi}{2}\right) + 2\beta \dot{\phi} \right]  \,,
%    \end{array}
    \label{oscillator-eq}
\end{eqnarray}

\noindent where 

\begin{eqnarray}
    \centering
%    \begin{array}{cl}
     \omega_0 = \dfrac{\gamma}{2M_s}\sqrt{\mathcal{K}_1\mathcal{K}_2} \ \hspace{0.25cm}{\text{and}}\hspace{0.25cm}
    \beta = \dfrac{\alpha\gamma}{4M_s}\left(\mathcal{K}_1 \lambda+\dfrac{\mathcal{K}_2}{\lambda}\right). 
%    \end{array}
    \label{wo-beta}
\end{eqnarray}

\noindent In this context, Eqs. \eqref{oscillator-eq-q} and \eqref{oscillator-eq} describe a pair of uncoupled harmonic oscillators with natural frequency $\omega_0$ and damping parameter $\beta$.

\begin{figure}[htbp]
    \centering
    \includegraphics[width=6.0cm ,angle=0]
    {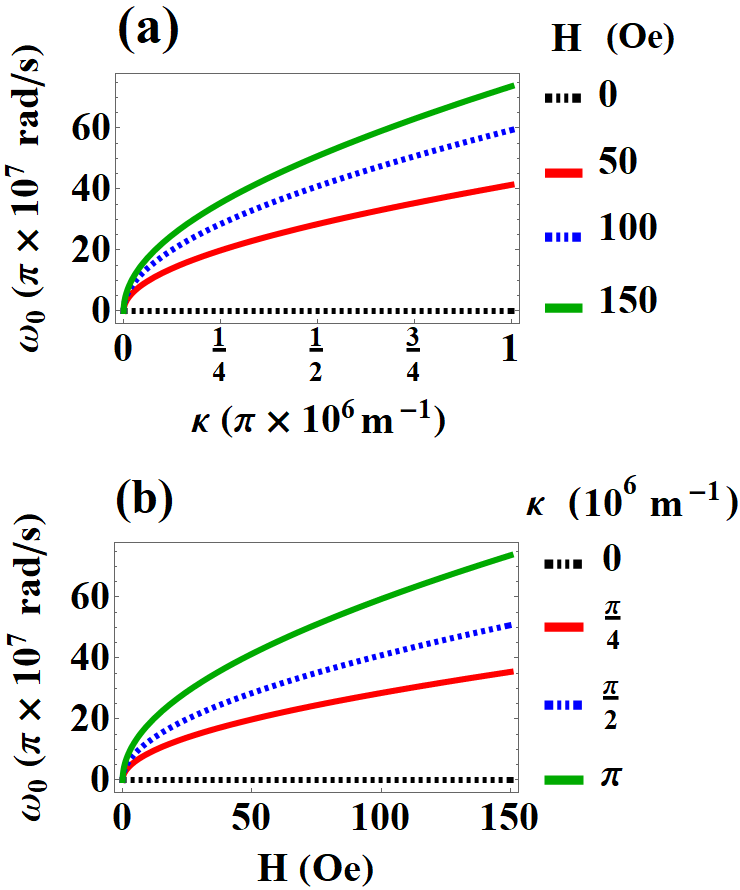}
    \caption{Frequency of the DW oscillation around the equilibrium position for \textbf{(a)} a fix curvature  $\kappa = \pi \times 10^6$ m$^{-1}$ and different $H$ values; and \textbf{(b)} a fix value of $H = 100$ Oe and different values of $\kappa$.}
    \label{FreqDamp}
    \end{figure}

\noindent Therefore, the above equations reveal that the DW oscillatory frequency can be controlled through the magnetic field and the NW curvature. Fig. \ref{FreqDamp} shows the behavior of $\omega_0$ as a function of $\kappa$ (Fig. \ref{FreqDamp}-a) and $H$ (\ref{FreqDamp}-b) for an NW with $\Delta r = 20$ nm and $\Delta z = 10$ nm. The increase of $\omega_0$ with $H$ and $\kappa$ corroborates the experimental results of Saitoh \textit{et al.}\cite{Saitoh2004}. Also, when we consider the limit of small curvature, our results agree with the predictions of Kr\"uger \textit{et al.} \cite{PhysRevB.75.054421}, who analyzed the DW dynamics in a curved structure through a theoretical model considering a straight NW in the presence of a magnetic field gradient. Additionally, the results presented in Eqs. \eqref{wo-beta} also points out the possibility of determining the damping constant using the described experimental setting \cite{Saitoh2004,PhysRevB.75.054421}.

\begin{figure}[htbp]
    \centering
    \includegraphics[width=4.5cm ,angle=0]
    {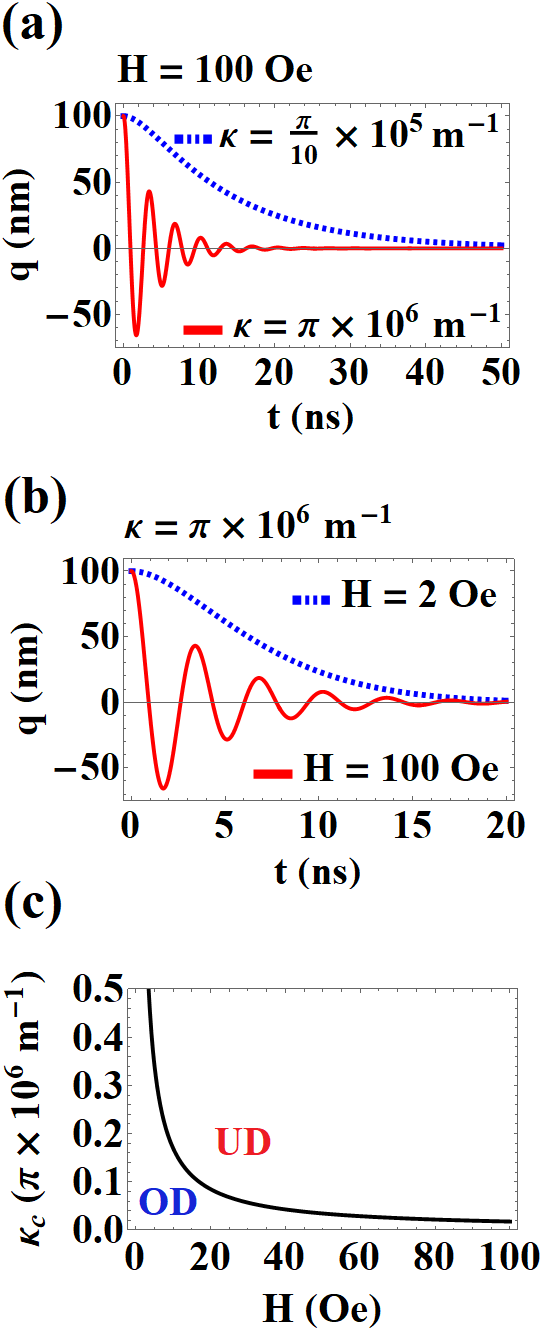}
    \caption{DW position $q$ as a function of time $t$ in the under-damped (red line) and over-damped (blue dashed line) regimes. \textbf{(a)} Considers a fix $H=100$ Oe and different $\kappa$ values.  \textbf{(b)}  Considers a fix $\kappa=\pi \times 10^6$ m$^{-1}$ and different $H$ values.  \textbf{(c)} Illustration of the critical curvature $\kappa_c$ that defines the transition between UD and OD regimes as a function of $H$.}
    \label{dynamics}
    \end{figure}

In this work, we focus on describing the oscillations of the DW position, but the phase oscillations show the same frequency. In this case, by disregarding $\alpha^2$, the solution of Eq. \eqref{oscillator-eq-q} is written as 

\begin{equation}
    \centering
%    \begin{array}{cl}
     q(t) = e^{-t\beta}\left( \mathcal{C}_1 e^{t\sqrt{\beta^2-\omega_0^2}} + \mathcal{C}_2 e^{-t\sqrt{\beta^2-\omega_0^2}}\right)\,, 
     \label{solution-harmonic}
\end{equation}

\noindent where the constants $\mathcal{C}_1$ and $\mathcal{C}_2$ are determined by the initial conditions. The obtained solution reveals that the DW motion presents two main behaviors depending on $\beta$ separated by the critical damping $\beta_c=\omega_0$. If $\beta<\beta_c$, the DW oscillates in an underdamped regime with oscillations whose amplitude gradually decreases to zero. On the other hand, for $\beta>\beta_c$, the DW oscillates in an overdamped regime, where $q(t)$ exponentially decays to the equilibrium position with no oscillations. Since $\beta$ depends on $H$ and $\kappa$, these parameters can be used to properly control the DW oscillation frequencies, as shown in Fig. \ref{dynamics}, where we present the DW position as a function of  time for different  NW curvatures (Fig. \ref{dynamics}-a) and magnetic field (Fig. \ref{dynamics}-b). Additionally, Fig. \ref{dynamics}-c presents the parameter regions where UD and OD regimes occur. We highlight that an equivalent behavior occurs for oscillations in the DW phase, which can be obtained by solving Eq. \eqref{oscillator-eq}. 

Finally, to compare our results with the expression obtained by Saitoh \textit{et al.} \cite{Saitoh2004} from mechanical analogy, we calculate the DW oscillation squared frequency $f_0^2 = \omega_0^2/(4\pi^2)$, which, in the limit  $\kappa \lambda \ll 1$, is given by

\begin{equation}
f_0^2 = \frac{ H M_s \kappa}{2\pi^2 m_{\mathcal{S}}} \, ,
\label{f0squared}
\end{equation}

\noindent 
where $m_\mathcal{S}=m/\mathcal{S}$ is the DW effective mass per unit area, derived using the analogy with of the harmonic oscillator for the DW position, {\it i.e.} $\omega_0=\sqrt{\mathcal{K}_1/m_{\mathcal{S}}}$. Therefore, the DW effective mass is evaluated as

\begin{equation}
    \centering
    m = \dfrac{\mathcal{S}}{\mathcal{K}_2}\left(\dfrac{2M_s}{\gamma}\right)^2 \, \, .
     \label{mass}
\end{equation}

\begin{figure}[!tbp]
    \centering
    \includegraphics[width=5cm ,angle=0]
    {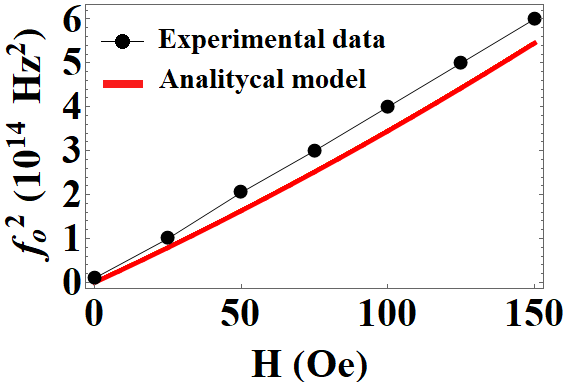}
    \caption{Frequency squared as a function of $H$ from the theoretical model (solid red line) and  experimental measurements in Ref. [\cite{Saitoh2004}] (black dots).  }
    \label{saitoh}
    \end{figure}

To compare results from our  theoretical model and the experimental results by Saitoh \textit{et al.}, we consider the equivalent NW with cross-section dimensions $\Delta r = 70$ nm, $\Delta z = 45$ nm, and $R= 50 \times 10 ^{-6}$ m.  Fig. \ref{saitoh} depicts results from Eq. \eqref{f0squared} (red line) and experimental data (black dots), showing a very good agreement. We also highlight that Saitoh \textit{et al.} estimated a DW effective mass of $m \approx 6 \times 10^{-23}$ kg at $H=100$ Oe. The theoretical model presented here predicts $m \in [7, 9] \times 10^{-23}$ kg, for $H \in [0, 150]$ Oe. 

In summary, using an approach based on the Lagrangian formalism, with an adequate definition of the coordinate system, it is possible to obtain the frequency of a DW oscillating along an NW. The obtained expressions indicate that such frequency can be  controlled by the magnetic field, the saturation magnetization, and the NW cross-section. Also, the DW dynamics can be reduced to a mechanical analogy of a pendulum under the action of a gravitational field. The mechanical analogy also allowed us to show that the DW oscillation regime can be over or underdamped, depending on the magnetic field and NW curvature. Finally, we could introduce an effective DW mass in bent nanowires, which agrees with experimental measurements.  These results could be used to  design nanowires with high-frequency responses for particular applications in DW-based spintronic devices.

We acknowledge financial support in Chile from FONDECYT Grant no. 1220215, and Financiamiento Basal para Centros Cient\'{i}ficos y Tecnol\'{o}gicos de Excelencia AFB220001. In Brazil, we thank the financial support of Capes (Finance Code 001), CNPq (Grant No. 305256/2022), and Fapemig (Grant No. APQ-00648-22). V.L.C-S acknowledges Universidad de Santiago de Chile and CEDENNA for hospitality. O.C.-F. also acknowledges the financial support  by the grants PID2019-108075RB-C31  funded by the Ministry of Science and Innovation of Spain MCIN/AEI/ 10.13039/501100011033. R.M acknowledges the postdoctoral fellowship program of Conicet Argentina.

\bibliographystyle{apsrev}
\bibliography{references}

\providecommand{\newblock}{}
\begin{thebibliography}{65}
\expandafter\ifx\csname natexlab\endcsname\relax\def\natexlab#1{#1}\fi
\expandafter\ifx\csname bibnamefont\endcsname\relax
  \def\bibnamefont#1{#1}\fi
\expandafter\ifx\csname bibfnamefont\endcsname\relax
  \def\bibfnamefont#1{#1}\fi
\expandafter\ifx\csname citenamefont\endcsname\relax
  \def\citenamefont#1{#1}\fi
\expandafter\ifx\csname url\endcsname\relax
  \def\url#1{\texttt{#1}}\fi
\expandafter\ifx\csname urlprefix\endcsname\relax\def\urlprefix{URL }\fi
\providecommand{\bibinfo}[2]{#2}
\providecommand{\eprint}[2][]{\url{#2}}

\bibitem[{\citenamefont{Moreno et~al.}(2021)\citenamefont{Moreno, Bran, Vazquez, and Kosel}}]{hilosMateriales}
\bibinfo{author}{\bibfnamefont{J.~A.} \bibnamefont{Moreno}}, \bibinfo{author}{\bibfnamefont{C.}~\bibnamefont{Bran}}, \bibinfo{author}{\bibfnamefont{M.}~\bibnamefont{Vazquez}}, \bibnamefont{and} \bibinfo{author}{\bibfnamefont{J.}~\bibnamefont{Kosel}}, \bibinfo{journal}{IEEE Transactions on Magnetics} \textbf{\bibinfo{volume}{57}}, \bibinfo{pages}{1} (\bibinfo{year}{2021}).

\bibitem[{\citenamefont{Cui et~al.}(2013)\citenamefont{Cui, Shi, Yuan, and Fu}}]{hilosAgua}
\bibinfo{author}{\bibfnamefont{H.-J.} \bibnamefont{Cui}}, \bibinfo{author}{\bibfnamefont{J.-W.} \bibnamefont{Shi}}, \bibinfo{author}{\bibfnamefont{B.}~\bibnamefont{Yuan}}, \bibnamefont{and} \bibinfo{author}{\bibfnamefont{M.-L.} \bibnamefont{Fu}}, \bibinfo{journal}{J. Mater. Chem. A} \textbf{\bibinfo{volume}{1}}, \bibinfo{pages}{5902} (\bibinfo{year}{2013}), \urlprefix\url{http://dx.doi.org/10.1039/C3TA01692G}.

\bibitem[{\citenamefont{Fernandez-Roldan et~al.}(2018)\citenamefont{Fernandez-Roldan, Serantes, del Real, Vazquez, and Chubykalo-Fesenko}}]{hiloshyperthermia}
\bibinfo{author}{\bibfnamefont{J.~A.} \bibnamefont{Fernandez-Roldan}}, \bibinfo{author}{\bibfnamefont{D.}~\bibnamefont{Serantes}}, \bibinfo{author}{\bibfnamefont{R.~P.} \bibnamefont{del Real}}, \bibinfo{author}{\bibfnamefont{M.}~\bibnamefont{Vazquez}}, \bibnamefont{and} \bibinfo{author}{\bibfnamefont{O.}~\bibnamefont{Chubykalo-Fesenko}}, \bibinfo{journal}{Applied Physics Letters} \textbf{\bibinfo{volume}{112}}, \bibinfo{pages}{212402} (\bibinfo{year}{2018}), \eprint{https://doi.org/10.1063/1.5025922}, \urlprefix\url{https://doi.org/10.1063/1.5025922}.

\bibitem[{\citenamefont{Muxworthy and Williams}(2015)}]{hilospalaeomagnetismo}
\bibinfo{author}{\bibfnamefont{A.}~\bibnamefont{Muxworthy}} \bibnamefont{and} \bibinfo{author}{\bibfnamefont{W.}~\bibnamefont{Williams}}, \bibinfo{journal}{Geophysical Journal International} \textbf{\bibinfo{volume}{202}}, \bibinfo{pages}{578} (\bibinfo{year}{2015}).

\bibitem[{\citenamefont{Ivanov et~al.}(2013)\citenamefont{Ivanov, V{\'{a}}zquez, and Chubykalo-Fesenko}}]{IvanovReversalModes}
\bibinfo{author}{\bibfnamefont{Y.~P.} \bibnamefont{Ivanov}}, \bibinfo{author}{\bibfnamefont{M.}~\bibnamefont{V{\'{a}}zquez}}, \bibnamefont{and} \bibinfo{author}{\bibfnamefont{O.}~\bibnamefont{Chubykalo-Fesenko}}, \bibinfo{journal}{J. Phys. D: Appl. Phys.} \textbf{\bibinfo{volume}{46}}, \bibinfo{pages}{485001} (\bibinfo{year}{2013}).

\bibitem[{\citenamefont{Moreno et~al.}(2022)\citenamefont{Moreno, Carvalho-Santos, Altbir, and Chubykalo-Fesenko}}]{MORENO2022168495}
\bibinfo{author}{\bibfnamefont{R.}~\bibnamefont{Moreno}}, \bibinfo{author}{\bibfnamefont{V.}~\bibnamefont{Carvalho-Santos}}, \bibinfo{author}{\bibfnamefont{D.}~\bibnamefont{Altbir}}, \bibnamefont{and} \bibinfo{author}{\bibfnamefont{O.}~\bibnamefont{Chubykalo-Fesenko}}, \bibinfo{journal}{J. Magn. Mag. Mat.} \textbf{\bibinfo{volume}{542}}, \bibinfo{pages}{168495} (\bibinfo{year}{2022}).

\bibitem[{\citenamefont{Parkin et~al.}(2008)\citenamefont{Parkin, Hayashi, and Thomas}}]{Parkin190}
\bibinfo{author}{\bibfnamefont{S.~S.~P.} \bibnamefont{Parkin}}, \bibinfo{author}{\bibfnamefont{M.}~\bibnamefont{Hayashi}}, \bibnamefont{and} \bibinfo{author}{\bibfnamefont{L.}~\bibnamefont{Thomas}}, \bibinfo{journal}{Science} \textbf{\bibinfo{volume}{320}}, \bibinfo{pages}{190} (\bibinfo{year}{2008}), ISSN \bibinfo{issn}{0036-8075}, \eprint{https://science.sciencemag.org/content/320/5873/190.full.pdf}, \urlprefix\url{https://science.sciencemag.org/content/320/5873/190}.

\bibitem[{\citenamefont{Sharma et~al.}(2015{\natexlab{a}})\citenamefont{Sharma, Muralidharan, and Tulapurkar}}]{Sharma2015}
\bibinfo{author}{\bibfnamefont{S.}~\bibnamefont{Sharma}}, \bibinfo{author}{\bibfnamefont{B.}~\bibnamefont{Muralidharan}}, \bibnamefont{and} \bibinfo{author}{\bibfnamefont{A.}~\bibnamefont{Tulapurkar}}, \bibinfo{journal}{Scientific Reports} \textbf{\bibinfo{volume}{5}}, \bibinfo{pages}{14647} (\bibinfo{year}{2015}{\natexlab{a}}), ISSN \bibinfo{issn}{2045-2322}, \urlprefix\url{https://doi.org/10.1038/srep14647}.

\bibitem[{\citenamefont{Nikonov et~al.}(2014)\citenamefont{Nikonov, Manipatruni, and Young}}]{doi:10.1063/1.4881061}
\bibinfo{author}{\bibfnamefont{D.~E.} \bibnamefont{Nikonov}}, \bibinfo{author}{\bibfnamefont{S.}~\bibnamefont{Manipatruni}}, \bibnamefont{and} \bibinfo{author}{\bibfnamefont{I.~A.} \bibnamefont{Young}}, \bibinfo{journal}{Journal of Applied Physics} \textbf{\bibinfo{volume}{115}}, \bibinfo{pages}{213902} (\bibinfo{year}{2014}), \eprint{https://doi.org/10.1063/1.4881061}, \urlprefix\url{https://doi.org/10.1063/1.4881061}.

\bibitem[{\citenamefont{Allwood et~al.}(2002)\citenamefont{Allwood, Xiong, Cooke, Faulkner, Atkinson, Vernier, and Cowburn}}]{Allwood2003}
\bibinfo{author}{\bibfnamefont{D.~A.} \bibnamefont{Allwood}}, \bibinfo{author}{\bibfnamefont{G.}~\bibnamefont{Xiong}}, \bibinfo{author}{\bibfnamefont{M.~D.} \bibnamefont{Cooke}}, \bibinfo{author}{\bibfnamefont{C.~C.} \bibnamefont{Faulkner}}, \bibinfo{author}{\bibfnamefont{D.}~\bibnamefont{Atkinson}}, \bibinfo{author}{\bibfnamefont{N.}~\bibnamefont{Vernier}}, \bibnamefont{and} \bibinfo{author}{\bibfnamefont{R.~P.} \bibnamefont{Cowburn}}, \bibinfo{journal}{Science} \textbf{\bibinfo{volume}{296}}, \bibinfo{pages}{2003} (\bibinfo{year}{2002}), ISSN \bibinfo{issn}{0036-8075}.

\bibitem[{\citenamefont{Espejo et~al.}(2017)\citenamefont{Espejo, Tejo~Lazo, Vidal, and Escrig}}]{Alvaro}
\bibinfo{author}{\bibfnamefont{A.}~\bibnamefont{Espejo}}, \bibinfo{author}{\bibfnamefont{F.}~\bibnamefont{Tejo~Lazo}}, \bibinfo{author}{\bibfnamefont{N.}~\bibnamefont{Vidal}}, \bibnamefont{and} \bibinfo{author}{\bibfnamefont{J.}~\bibnamefont{Escrig}}, \bibinfo{journal}{Sci. Rep.} \textbf{\bibinfo{volume}{7}}, \bibinfo{pages}{4736} (\bibinfo{year}{2017}).

\bibitem[{\citenamefont{Voto and Lopez-Diaz}(2017)}]{Voto}
\bibinfo{author}{\bibfnamefont{M.}~\bibnamefont{Voto}} \bibnamefont{and} \bibinfo{author}{\bibfnamefont{L.}~\bibnamefont{Lopez-Diaz}}, \bibinfo{journal}{Scientific Reports} \textbf{\bibinfo{volume}{7}}, \bibinfo{pages}{13559} (\bibinfo{year}{2017}).

\bibitem[{\citenamefont{Berger}(1984)}]{Berger}
\bibinfo{author}{\bibfnamefont{L.}~\bibnamefont{Berger}}, \bibinfo{journal}{Journal of Applied Physics} \textbf{\bibinfo{volume}{55}}, \bibinfo{pages}{1954} (\bibinfo{year}{1984}), \eprint{https://doi.org/10.1063/1.333530}, \urlprefix\url{https://doi.org/10.1063/1.333530}.

\bibitem[{\citenamefont{Hung and Berger}(1988)}]{Hung}
\bibinfo{author}{\bibfnamefont{C.}~\bibnamefont{Hung}} \bibnamefont{and} \bibinfo{author}{\bibfnamefont{L.}~\bibnamefont{Berger}}, \bibinfo{journal}{Journal of Applied Physics} \textbf{\bibinfo{volume}{63}}, \bibinfo{pages}{4276} (\bibinfo{year}{1988}), \eprint{https://doi.org/10.1063/1.340201}, \urlprefix\url{https://doi.org/10.1063/1.340201}.

\bibitem[{\citenamefont{Yamanouchi et~al.}(2004)\citenamefont{Yamanouchi, Chiba, Matsukura, and Ohno}}]{Yamanouchi2004}
\bibinfo{author}{\bibfnamefont{M.}~\bibnamefont{Yamanouchi}}, \bibinfo{author}{\bibfnamefont{D.}~\bibnamefont{Chiba}}, \bibinfo{author}{\bibfnamefont{F.}~\bibnamefont{Matsukura}}, \bibnamefont{and} \bibinfo{author}{\bibfnamefont{H.}~\bibnamefont{Ohno}}, \bibinfo{journal}{Nature} \textbf{\bibinfo{volume}{428}}, \bibinfo{pages}{539} (\bibinfo{year}{2004}), ISSN \bibinfo{issn}{1476-4687}, \urlprefix\url{https://doi.org/10.1038/nature02441}.

\bibitem[{\citenamefont{Zhang and Li}(2004)}]{PhysRevLett.93.127204}
\bibinfo{author}{\bibfnamefont{S.}~\bibnamefont{Zhang}} \bibnamefont{and} \bibinfo{author}{\bibfnamefont{Z.}~\bibnamefont{Li}}, \bibinfo{journal}{Phys. Rev. Lett.} \textbf{\bibinfo{volume}{93}}, \bibinfo{pages}{127204} (\bibinfo{year}{2004}), \urlprefix\url{https://link.aps.org/doi/10.1103/PhysRevLett.93.127204}.

\bibitem[{\citenamefont{Albrecht et~al.}(2005)\citenamefont{Albrecht, Hu, Guhr, Ulbrich, Boneberg, Leiderer, and Schatz}}]{ExpAlbrecht2005}
\bibinfo{author}{\bibfnamefont{M.}~\bibnamefont{Albrecht}}, \bibinfo{author}{\bibfnamefont{G.}~\bibnamefont{Hu}}, \bibinfo{author}{\bibfnamefont{I.~L.} \bibnamefont{Guhr}}, \bibinfo{author}{\bibfnamefont{T.~C.} \bibnamefont{Ulbrich}}, \bibinfo{author}{\bibfnamefont{J.}~\bibnamefont{Boneberg}}, \bibinfo{author}{\bibfnamefont{P.}~\bibnamefont{Leiderer}}, \bibnamefont{and} \bibinfo{author}{\bibfnamefont{G.}~\bibnamefont{Schatz}}, \bibinfo{journal}{Nature Materials} \textbf{\bibinfo{volume}{4}}, \bibinfo{pages}{203} (\bibinfo{year}{2005}), ISSN \bibinfo{issn}{1476-4660}, \urlprefix\url{https://doi.org/10.1038/nmat1324}.

\bibitem[{\citenamefont{Minguez-Bacho et~al.}(2014)\citenamefont{Minguez-Bacho, Rodriguez-López, Vázquez, Hernández-Vélez, and Nielsch}}]{ExpMinguez-Bacho_2014}
\bibinfo{author}{\bibfnamefont{I.}~\bibnamefont{Minguez-Bacho}}, \bibinfo{author}{\bibfnamefont{S.}~\bibnamefont{Rodriguez-López}}, \bibinfo{author}{\bibfnamefont{M.}~\bibnamefont{Vázquez}}, \bibinfo{author}{\bibfnamefont{M.}~\bibnamefont{Hernández-Vélez}}, \bibnamefont{and} \bibinfo{author}{\bibfnamefont{K.}~\bibnamefont{Nielsch}}, \bibinfo{journal}{Nanotechnology} \textbf{\bibinfo{volume}{25}}, \bibinfo{pages}{145301} (\bibinfo{year}{2014}), \urlprefix\url{https://dx.doi.org/10.1088/0957-4484/25/14/145301}.

\bibitem[{\citenamefont{Schmidt and Eberl}(2001)}]{ExpRollingTubes}
\bibinfo{author}{\bibfnamefont{O.~G.} \bibnamefont{Schmidt}} \bibnamefont{and} \bibinfo{author}{\bibfnamefont{K.}~\bibnamefont{Eberl}}, \bibinfo{journal}{Nature} \textbf{\bibinfo{volume}{410}}, \bibinfo{pages}{168} (\bibinfo{year}{2001}), ISSN \bibinfo{issn}{1476-4687}, \urlprefix\url{https://doi.org/10.1038/35065525}.

\bibitem[{\citenamefont{Streubel et~al.}(2012)\citenamefont{Streubel, Kravchuk, Sheka, Makarov, Kronast, Schmidt, and Gaididei}}]{ExpSphericalCaps}
\bibinfo{author}{\bibfnamefont{R.}~\bibnamefont{Streubel}}, \bibinfo{author}{\bibfnamefont{V.~P.} \bibnamefont{Kravchuk}}, \bibinfo{author}{\bibfnamefont{D.~D.} \bibnamefont{Sheka}}, \bibinfo{author}{\bibfnamefont{D.}~\bibnamefont{Makarov}}, \bibinfo{author}{\bibfnamefont{F.}~\bibnamefont{Kronast}}, \bibinfo{author}{\bibfnamefont{O.~G.} \bibnamefont{Schmidt}}, \bibnamefont{and} \bibinfo{author}{\bibfnamefont{Y.}~\bibnamefont{Gaididei}}, \bibinfo{journal}{Applied Physics Letters} \textbf{\bibinfo{volume}{101}}, \bibinfo{pages}{132419} (\bibinfo{year}{2012}), \eprint{https://doi.org/10.1063/1.4756708}, \urlprefix\url{https://doi.org/10.1063/1.4756708}.

\bibitem[{\citenamefont{Tejo et~al.}(2020)\citenamefont{Tejo, Toneto, Oyarz{\'u}n, Hermosilla, Danna, Palma, da~Silva, Dorneles, and Denardin}}]{ExpTejo2020}
\bibinfo{author}{\bibfnamefont{F.}~\bibnamefont{Tejo}}, \bibinfo{author}{\bibfnamefont{D.}~\bibnamefont{Toneto}}, \bibinfo{author}{\bibfnamefont{S.}~\bibnamefont{Oyarz{\'u}n}}, \bibinfo{author}{\bibfnamefont{J.}~\bibnamefont{Hermosilla}}, \bibinfo{author}{\bibfnamefont{C.~S.} \bibnamefont{Danna}}, \bibinfo{author}{\bibfnamefont{J.~L.} \bibnamefont{Palma}}, \bibinfo{author}{\bibfnamefont{R.~B.} \bibnamefont{da~Silva}}, \bibinfo{author}{\bibfnamefont{L.~S.} \bibnamefont{Dorneles}}, \bibnamefont{and} \bibinfo{author}{\bibfnamefont{J.~C.} \bibnamefont{Denardin}}, \bibinfo{journal}{ACS Applied Materials {\&} Interfaces} \textbf{\bibinfo{volume}{12}}, \bibinfo{pages}{53454} (\bibinfo{year}{2020}), ISSN \bibinfo{issn}{1944-8244}, \urlprefix\url{https://doi.org/10.1021/acsami.0c14350}.

\bibitem[{\citenamefont{Sheka}(2021)}]{Ref1}
\bibinfo{author}{\bibfnamefont{D.~D.} \bibnamefont{Sheka}}, \bibinfo{journal}{Appl. Phys. Lett.} \textbf{\bibinfo{volume}{118}}, \bibinfo{pages}{230502} (\bibinfo{year}{2021}).

\bibitem[{\citenamefont{Makarov et~al.}(2022)\citenamefont{Makarov, Volkov, K\'akay, Pylypovskyi, Budinsk'a, and Dobrovolskiy}}]{Ref2}
\bibinfo{author}{\bibfnamefont{D.}~\bibnamefont{Makarov}}, \bibinfo{author}{\bibfnamefont{O.~M.~V.} \bibnamefont{Volkov}}, \bibinfo{author}{\bibfnamefont{A.}~\bibnamefont{K\'akay}}, \bibinfo{author}{\bibfnamefont{O.~V.} \bibnamefont{Pylypovskyi}}, \bibinfo{author}{\bibfnamefont{B.}~\bibnamefont{Budinsk'a}}, \bibnamefont{and} \bibinfo{author}{\bibfnamefont{O.~V.} \bibnamefont{Dobrovolskiy}}, \bibinfo{journal}{Adv. Mat.} \textbf{\bibinfo{volume}{34}}, \bibinfo{pages}{2101758} (\bibinfo{year}{2022}).

\bibitem[{\citenamefont{Yershov and Volkov}(2022)}]{Ref3}
\bibinfo{author}{\bibfnamefont{K.~V.} \bibnamefont{Yershov}} \bibnamefont{and} \bibinfo{author}{\bibfnamefont{O.~M.} \bibnamefont{Volkov}}, \emph{\bibinfo{title}{Geometry-Induced Magnetic Effects in Planar Curvilinear Nanosystems}} (\bibinfo{publisher}{Springer International Publishing}, \bibinfo{address}{Cham}, \bibinfo{year}{2022}), pp. \bibinfo{pages}{1--35}, ISBN \bibinfo{isbn}{978-3-031-09086-8}, \urlprefix\url{https://doi.org/10.1007/978-3-031-09086-8_1}.

\bibitem[{\citenamefont{Pylypovskyi et~al.}(2022)\citenamefont{Pylypovskyi, Phatak, and Volkov}}]{Ref4}
\bibinfo{author}{\bibfnamefont{O.}~\bibnamefont{Pylypovskyi}}, \bibinfo{author}{\bibfnamefont{C.}~\bibnamefont{Phatak}}, \bibnamefont{and} \bibinfo{author}{\bibfnamefont{O.}~\bibnamefont{Volkov}}, \emph{\bibinfo{title}{Effects of Curvature and Torsion on Magnetic Nanowires}} (\bibinfo{publisher}{Springer International Publishing}, \bibinfo{address}{Cham}, \bibinfo{year}{2022}), pp. \bibinfo{pages}{37--81}, ISBN \bibinfo{isbn}{978-3-031-09086-8}, \urlprefix\url{https://doi.org/10.1007/978-3-031-09086-8_2}.

\bibitem[{\citenamefont{Dobrovolskiy et~al.}(2022)\citenamefont{Dobrovolskiy, Pylypovskyi, Skoric, Fern{\'a}ndez-Pacheco, Van Den~Berg, Ladak, and Huth}}]{Ref5}
\bibinfo{author}{\bibfnamefont{O.~V.} \bibnamefont{Dobrovolskiy}}, \bibinfo{author}{\bibfnamefont{O.~V.} \bibnamefont{Pylypovskyi}}, \bibinfo{author}{\bibfnamefont{L.}~\bibnamefont{Skoric}}, \bibinfo{author}{\bibfnamefont{A.}~\bibnamefont{Fern{\'a}ndez-Pacheco}}, \bibinfo{author}{\bibfnamefont{A.}~\bibnamefont{Van Den~Berg}}, \bibinfo{author}{\bibfnamefont{S.}~\bibnamefont{Ladak}}, \bibnamefont{and} \bibinfo{author}{\bibfnamefont{M.}~\bibnamefont{Huth}}, \emph{\bibinfo{title}{Complex-Shaped 3D Nanoarchitectures for Magnetism and Superconductivity}} (\bibinfo{publisher}{Springer International Publishing}, \bibinfo{address}{Cham}, \bibinfo{year}{2022}), pp. \bibinfo{pages}{215--268}, ISBN \bibinfo{isbn}{978-3-031-09086-8}, \urlprefix\url{https://doi.org/10.1007/978-3-031-09086-8_5}.

\bibitem[{\citenamefont{Fernández-Pacheco et~al.}(2020)\citenamefont{Fernández-Pacheco, Skoric, De~Teresa, Pablo-Navarro, Huth, and Dobrovolskiy}}]{Pachecho2020}
\bibinfo{author}{\bibfnamefont{A.}~\bibnamefont{Fernández-Pacheco}}, \bibinfo{author}{\bibfnamefont{L.}~\bibnamefont{Skoric}}, \bibinfo{author}{\bibfnamefont{J.~M.} \bibnamefont{De~Teresa}}, \bibinfo{author}{\bibfnamefont{J.}~\bibnamefont{Pablo-Navarro}}, \bibinfo{author}{\bibfnamefont{M.}~\bibnamefont{Huth}}, \bibnamefont{and} \bibinfo{author}{\bibfnamefont{O.~V.} \bibnamefont{Dobrovolskiy}}, \bibinfo{journal}{Materials} \textbf{\bibinfo{volume}{13}}, \bibinfo{pages}{3774} (\bibinfo{year}{2020}), ISSN \bibinfo{issn}{1996-1944}, \urlprefix\url{http://dx.doi.org/10.3390/ma13173774}.

\bibitem[{\citenamefont{Zhang et~al.}(2021)\citenamefont{Zhang, Li, Jiang, Tang, Xu, Zhao, Fu, Zhou, and Chen}}]{helices}
\bibinfo{author}{\bibfnamefont{C.}~\bibnamefont{Zhang}}, \bibinfo{author}{\bibfnamefont{X.}~\bibnamefont{Li}}, \bibinfo{author}{\bibfnamefont{L.}~\bibnamefont{Jiang}}, \bibinfo{author}{\bibfnamefont{D.}~\bibnamefont{Tang}}, \bibinfo{author}{\bibfnamefont{H.}~\bibnamefont{Xu}}, \bibinfo{author}{\bibfnamefont{P.}~\bibnamefont{Zhao}}, \bibinfo{author}{\bibfnamefont{J.}~\bibnamefont{Fu}}, \bibinfo{author}{\bibfnamefont{Q.}~\bibnamefont{Zhou}}, \bibnamefont{and} \bibinfo{author}{\bibfnamefont{Y.}~\bibnamefont{Chen}}, \bibinfo{journal}{Advanced Functional Materials} \textbf{\bibinfo{volume}{31}}, \bibinfo{pages}{2102777} (\bibinfo{year}{2021}), \urlprefix\url{10.1002/adfm.202102777}.

\bibitem[{\citenamefont{Sanz-Hernández and Fernández-Pacheco}(2020)}]{SANZHERNANDEZ202085}
\bibinfo{author}{\bibfnamefont{D.}~\bibnamefont{Sanz-Hernández}} \bibnamefont{and} \bibinfo{author}{\bibfnamefont{A.}~\bibnamefont{Fernández-Pacheco}}, in \emph{\bibinfo{booktitle}{Magnetic Nano- and Microwires (Second Edition)}}, edited by \bibinfo{editor}{\bibfnamefont{M.}~\bibnamefont{Vázquez}} (\bibinfo{publisher}{Woodhead Publishing}, \bibinfo{year}{2020}), Woodhead Publishing Series in Electronic and Optical Materials, pp. \bibinfo{pages}{85--102}, \bibinfo{edition}{second edition} ed., ISBN \bibinfo{isbn}{978-0-08-102832-2}, \urlprefix\url{https://www.sciencedirect.com/science/article/pii/B9780081028322000049}.

\bibitem[{\citenamefont{Sanz-Hern{\'a}ndez et~al.}(2020)\citenamefont{Sanz-Hern{\'a}ndez, Hierro-Rodriguez, Donnelly, Pablo-Navarro, Sorrentino, Pereiro, Mag{\'e}n, McVitie, de~Teresa, Ferrer et~al.}}]{Sanz-Hernández2020}
\bibinfo{author}{\bibfnamefont{D.}~\bibnamefont{Sanz-Hern{\'a}ndez}}, \bibinfo{author}{\bibfnamefont{A.}~\bibnamefont{Hierro-Rodriguez}}, \bibinfo{author}{\bibfnamefont{C.}~\bibnamefont{Donnelly}}, \bibinfo{author}{\bibfnamefont{J.}~\bibnamefont{Pablo-Navarro}}, \bibinfo{author}{\bibfnamefont{A.}~\bibnamefont{Sorrentino}}, \bibinfo{author}{\bibfnamefont{E.}~\bibnamefont{Pereiro}}, \bibinfo{author}{\bibfnamefont{C.}~\bibnamefont{Mag{\'e}n}}, \bibinfo{author}{\bibfnamefont{S.}~\bibnamefont{McVitie}}, \bibinfo{author}{\bibfnamefont{J.~M.} \bibnamefont{de~Teresa}}, \bibinfo{author}{\bibfnamefont{S.}~\bibnamefont{Ferrer}}, \bibnamefont{et~al.}, \bibinfo{journal}{ACS Nano} \textbf{\bibinfo{volume}{14}}, \bibinfo{pages}{8084} (\bibinfo{year}{2020}), ISSN \bibinfo{issn}{1936-0851}, \urlprefix\url{https://doi.org/10.1021/acsnano.0c00720}.

\bibitem[{\citenamefont{Fern{\'a}ndez-Pacheco et~al.}(2013)\citenamefont{Fern{\'a}ndez-Pacheco, Serrano-Ram{\'o}n, Michalik, Ibarra, De~Teresa, O'Brien, Petit, Lee, and Cowburn}}]{ExpFernández-Pacheco2013}
\bibinfo{author}{\bibfnamefont{A.}~\bibnamefont{Fern{\'a}ndez-Pacheco}}, \bibinfo{author}{\bibfnamefont{L.}~\bibnamefont{Serrano-Ram{\'o}n}}, \bibinfo{author}{\bibfnamefont{J.~M.} \bibnamefont{Michalik}}, \bibinfo{author}{\bibfnamefont{M.~R.} \bibnamefont{Ibarra}}, \bibinfo{author}{\bibfnamefont{J.~M.} \bibnamefont{De~Teresa}}, \bibinfo{author}{\bibfnamefont{L.}~\bibnamefont{O'Brien}}, \bibinfo{author}{\bibfnamefont{D.}~\bibnamefont{Petit}}, \bibinfo{author}{\bibfnamefont{J.}~\bibnamefont{Lee}}, \bibnamefont{and} \bibinfo{author}{\bibfnamefont{R.~P.} \bibnamefont{Cowburn}}, \bibinfo{journal}{Scientific Reports} \textbf{\bibinfo{volume}{3}}, \bibinfo{pages}{1492} (\bibinfo{year}{2013}), ISSN \bibinfo{issn}{2045-2322}, \urlprefix\url{https://doi.org/10.1038/srep01492}.

\bibitem[{\citenamefont{Skoric et~al.}(2022)\citenamefont{Skoric, Donnelly, Hierro-Rodriguez, Sandoval, Ruiz-Gomez, Foerster, Niño, Belkhou, Abert, Suess et~al.}}]{locomotion}
\bibinfo{author}{\bibfnamefont{L.}~\bibnamefont{Skoric}}, \bibinfo{author}{\bibfnamefont{C.}~\bibnamefont{Donnelly}}, \bibinfo{author}{\bibfnamefont{A.}~\bibnamefont{Hierro-Rodriguez}}, \bibinfo{author}{\bibfnamefont{M.}~\bibnamefont{Sandoval}}, \bibinfo{author}{\bibfnamefont{S.}~\bibnamefont{Ruiz-Gomez}}, \bibinfo{author}{\bibfnamefont{M.}~\bibnamefont{Foerster}}, \bibinfo{author}{\bibfnamefont{M.}~\bibnamefont{Niño}}, \bibinfo{author}{\bibfnamefont{R.}~\bibnamefont{Belkhou}}, \bibinfo{author}{\bibfnamefont{C.}~\bibnamefont{Abert}}, \bibinfo{author}{\bibfnamefont{D.}~\bibnamefont{Suess}}, \bibnamefont{et~al.}, \bibinfo{journal}{ACS Nano} \textbf{\bibinfo{volume}{16}} (\bibinfo{year}{2022}).

\bibitem[{\citenamefont{Altbir et~al.}(2020)\citenamefont{Altbir, Fonseca, Chubykalo-Fesenko, Corona, Moreno, Carvalho-Santos, and Ivanov}}]{Altbir2020}
\bibinfo{author}{\bibfnamefont{D.}~\bibnamefont{Altbir}}, \bibinfo{author}{\bibfnamefont{J.~M.} \bibnamefont{Fonseca}}, \bibinfo{author}{\bibfnamefont{O.}~\bibnamefont{Chubykalo-Fesenko}}, \bibinfo{author}{\bibfnamefont{R.~M.} \bibnamefont{Corona}}, \bibinfo{author}{\bibfnamefont{R.}~\bibnamefont{Moreno}}, \bibinfo{author}{\bibfnamefont{V.~L.} \bibnamefont{Carvalho-Santos}}, \bibnamefont{and} \bibinfo{author}{\bibfnamefont{Y.~P.} \bibnamefont{Ivanov}}, \bibinfo{journal}{Sci. Rep.} \textbf{\bibinfo{volume}{10}}, \bibinfo{pages}{21911} (\bibinfo{year}{2020}), ISSN \bibinfo{issn}{2045-2322}.

\bibitem[{\citenamefont{Moreno et~al.}(2017)\citenamefont{Moreno, Carvalho-Santos, Espejo, Laroze, Chubykalo-Fesenko, and Altbir}}]{PhysRevB.96.184401}
\bibinfo{author}{\bibfnamefont{R.}~\bibnamefont{Moreno}}, \bibinfo{author}{\bibfnamefont{V.~L.} \bibnamefont{Carvalho-Santos}}, \bibinfo{author}{\bibfnamefont{A.~P.} \bibnamefont{Espejo}}, \bibinfo{author}{\bibfnamefont{D.}~\bibnamefont{Laroze}}, \bibinfo{author}{\bibfnamefont{O.}~\bibnamefont{Chubykalo-Fesenko}}, \bibnamefont{and} \bibinfo{author}{\bibfnamefont{D.}~\bibnamefont{Altbir}}, \bibinfo{journal}{Phys. Rev. B} \textbf{\bibinfo{volume}{96}}, \bibinfo{pages}{184401} (\bibinfo{year}{2017}).

\bibitem[{\citenamefont{Bittencourt et~al.}(2021)\citenamefont{Bittencourt, Moreno, Cacilhas, Castillo-Sepúlveda, Chubykalo-Fesenko, Altbir, and Carvalho-Santos}}]{secondWalker}
\bibinfo{author}{\bibfnamefont{G.~H.~R.} \bibnamefont{Bittencourt}}, \bibinfo{author}{\bibfnamefont{R.}~\bibnamefont{Moreno}}, \bibinfo{author}{\bibfnamefont{R.}~\bibnamefont{Cacilhas}}, \bibinfo{author}{\bibfnamefont{S.}~\bibnamefont{Castillo-Sepúlveda}}, \bibinfo{author}{\bibfnamefont{O.}~\bibnamefont{Chubykalo-Fesenko}}, \bibinfo{author}{\bibfnamefont{D.}~\bibnamefont{Altbir}}, \bibnamefont{and} \bibinfo{author}{\bibfnamefont{V.~L.} \bibnamefont{Carvalho-Santos}}, \bibinfo{journal}{Appl. Phys. Lett.} \textbf{\bibinfo{volume}{118}}, \bibinfo{pages}{142405} (\bibinfo{year}{2021}).

\bibitem[{\citenamefont{Yershov et~al.}(2016)\citenamefont{Yershov, Kravchuk, Sheka, and Gaididei}}]{Yershov-Helice}
\bibinfo{author}{\bibfnamefont{K.~V.} \bibnamefont{Yershov}}, \bibinfo{author}{\bibfnamefont{V.~P.} \bibnamefont{Kravchuk}}, \bibinfo{author}{\bibfnamefont{D.~D.} \bibnamefont{Sheka}}, \bibnamefont{and} \bibinfo{author}{\bibfnamefont{Y.}~\bibnamefont{Gaididei}}, \bibinfo{journal}{Phys. Rev. B} \textbf{\bibinfo{volume}{93}}, \bibinfo{pages}{094418} (\bibinfo{year}{2016}).

\bibitem[{\citenamefont{Porter and Donahue}(2004)}]{OsciStripes}
\bibinfo{author}{\bibfnamefont{D.~G.} \bibnamefont{Porter}} \bibnamefont{and} \bibinfo{author}{\bibfnamefont{M.~J.} \bibnamefont{Donahue}}, \bibinfo{journal}{J. Appl. Phys.} \textbf{\bibinfo{volume}{95}}, \bibinfo{pages}{6729} (\bibinfo{year}{2004}).

\bibitem[{\citenamefont{Schryer and Walker}(1974)}]{Walker}
\bibinfo{author}{\bibfnamefont{N.~L.} \bibnamefont{Schryer}} \bibnamefont{and} \bibinfo{author}{\bibfnamefont{L.~R.} \bibnamefont{Walker}}, \bibinfo{journal}{J. Appl. Phys.} \textbf{\bibinfo{volume}{45}}, \bibinfo{pages}{5406} (\bibinfo{year}{1974}).

\bibitem[{\citenamefont{Mougin et~al.}(2007)\citenamefont{Mougin, Cormier, Adam, Metaxas, and Ferré}}]{Mougin_2007}
\bibinfo{author}{\bibfnamefont{A.}~\bibnamefont{Mougin}}, \bibinfo{author}{\bibfnamefont{M.}~\bibnamefont{Cormier}}, \bibinfo{author}{\bibfnamefont{J.~P.} \bibnamefont{Adam}}, \bibinfo{author}{\bibfnamefont{P.~J.} \bibnamefont{Metaxas}}, \bibnamefont{and} \bibinfo{author}{\bibfnamefont{J.}~\bibnamefont{Ferré}}, \bibinfo{journal}{Europhys. Lett.} \textbf{\bibinfo{volume}{78}}, \bibinfo{pages}{57007} (\bibinfo{year}{2007}).

\bibitem[{\citenamefont{Yan et~al.}(2010)\citenamefont{Yan, Kakay, Gliga, and Hertel}}]{Hertel}
\bibinfo{author}{\bibfnamefont{M.}~\bibnamefont{Yan}}, \bibinfo{author}{\bibfnamefont{A.}~\bibnamefont{Kakay}}, \bibinfo{author}{\bibfnamefont{S.}~\bibnamefont{Gliga}}, \bibnamefont{and} \bibinfo{author}{\bibfnamefont{R.}~\bibnamefont{Hertel}}, \bibinfo{journal}{Phys. Rev. Lett.} \textbf{\bibinfo{volume}{104}}, \bibinfo{pages}{057201} (\bibinfo{year}{2010}).

\bibitem[{\citenamefont{Cacilhas et~al.}(2020)\citenamefont{Cacilhas, de~Araujo, Carvalho-Santos, Moreno, Chubykalo-Fesenko, and Altbir}}]{WalkerhilosCurvos}
\bibinfo{author}{\bibfnamefont{R.}~\bibnamefont{Cacilhas}}, \bibinfo{author}{\bibfnamefont{C.~I.~L.} \bibnamefont{de~Araujo}}, \bibinfo{author}{\bibfnamefont{V.~L.} \bibnamefont{Carvalho-Santos}}, \bibinfo{author}{\bibfnamefont{R.}~\bibnamefont{Moreno}}, \bibinfo{author}{\bibfnamefont{O.}~\bibnamefont{Chubykalo-Fesenko}}, \bibnamefont{and} \bibinfo{author}{\bibfnamefont{D.}~\bibnamefont{Altbir}}, \bibinfo{journal}{Phys. Rev. B} \textbf{\bibinfo{volume}{101}}, \bibinfo{pages}{184418} (\bibinfo{year}{2020}).

\bibitem[{\citenamefont{Caretta et~al.}(2020)\citenamefont{Caretta, Oh, Fakhrul, Lee, Lee, Kim, Ross, Lee, and Beach}}]{relativity1}
\bibinfo{author}{\bibfnamefont{L.}~\bibnamefont{Caretta}}, \bibinfo{author}{\bibfnamefont{S.-H.} \bibnamefont{Oh}}, \bibinfo{author}{\bibfnamefont{T.}~\bibnamefont{Fakhrul}}, \bibinfo{author}{\bibfnamefont{D.-K.} \bibnamefont{Lee}}, \bibinfo{author}{\bibfnamefont{B.~H.} \bibnamefont{Lee}}, \bibinfo{author}{\bibfnamefont{S.~K.} \bibnamefont{Kim}}, \bibinfo{author}{\bibfnamefont{C.~A.} \bibnamefont{Ross}}, \bibinfo{author}{\bibfnamefont{K.-J.} \bibnamefont{Lee}}, \bibnamefont{and} \bibinfo{author}{\bibfnamefont{G.~S.~D.} \bibnamefont{Beach}}, \bibinfo{journal}{Science} \textbf{\bibinfo{volume}{370}}, \bibinfo{pages}{1438} (\bibinfo{year}{2020}).

\bibitem[{\citenamefont{Yan et~al.}(2013)\citenamefont{Yan, K\'akay, Andreas, and Hertel}}]{Cherenkov}
\bibinfo{author}{\bibfnamefont{M.}~\bibnamefont{Yan}}, \bibinfo{author}{\bibfnamefont{A.}~\bibnamefont{K\'akay}}, \bibinfo{author}{\bibfnamefont{C.}~\bibnamefont{Andreas}}, \bibnamefont{and} \bibinfo{author}{\bibfnamefont{R.}~\bibnamefont{Hertel}}, \bibinfo{journal}{Phys. Rev. B} \textbf{\bibinfo{volume}{88}}, \bibinfo{pages}{220412} (\bibinfo{year}{2013}), \urlprefix\url{https://link.aps.org/doi/10.1103/PhysRevB.88.220412}.

\bibitem[{\citenamefont{Bittencourt et~al.}(2022{\natexlab{a}})\citenamefont{Bittencourt, Chubykalo-Fesenko, Altbir, Carvalho-Santos, and Moreno}}]{AreaLaw}
\bibinfo{author}{\bibfnamefont{G.~H.~R.} \bibnamefont{Bittencourt}}, \bibinfo{author}{\bibfnamefont{O.}~\bibnamefont{Chubykalo-Fesenko}}, \bibinfo{author}{\bibfnamefont{D.}~\bibnamefont{Altbir}}, \bibinfo{author}{\bibfnamefont{V.~L.} \bibnamefont{Carvalho-Santos}}, \bibnamefont{and} \bibinfo{author}{\bibfnamefont{R.}~\bibnamefont{Moreno}}, \bibinfo{journal}{Phys. Rev. B} \textbf{\bibinfo{volume}{106}}, \bibinfo{pages}{094410} (\bibinfo{year}{2022}{\natexlab{a}}), \urlprefix\url{https://link.aps.org/doi/10.1103/PhysRevB.106.094410}.

\bibitem[{\citenamefont{Volkov et~al.}(2019)\citenamefont{Volkov, K\'akay, Kronast, M\"onch, Mawass, Fassbender, and Makarov}}]{Volkov-PRL}
\bibinfo{author}{\bibfnamefont{O.~M.} \bibnamefont{Volkov}}, \bibinfo{author}{\bibfnamefont{A.}~\bibnamefont{K\'akay}}, \bibinfo{author}{\bibfnamefont{F.}~\bibnamefont{Kronast}}, \bibinfo{author}{\bibfnamefont{I.}~\bibnamefont{M\"onch}}, \bibinfo{author}{\bibfnamefont{M.-A.} \bibnamefont{Mawass}}, \bibinfo{author}{\bibfnamefont{J.}~\bibnamefont{Fassbender}}, \bibnamefont{and} \bibinfo{author}{\bibfnamefont{D.}~\bibnamefont{Makarov}}, \bibinfo{journal}{Phys. Rev. Lett.} \textbf{\bibinfo{volume}{123}}, \bibinfo{pages}{077201} (\bibinfo{year}{2019}).

\bibitem[{\citenamefont{Lewis et~al.}(2009)\citenamefont{Lewis, Petit, Thevenard, Jausovec, O’Brien, Read, and Cowburn}}]{Lewis-APL}
\bibinfo{author}{\bibfnamefont{E.~R.} \bibnamefont{Lewis}}, \bibinfo{author}{\bibfnamefont{D.}~\bibnamefont{Petit}}, \bibinfo{author}{\bibfnamefont{L.}~\bibnamefont{Thevenard}}, \bibinfo{author}{\bibfnamefont{A.~V.} \bibnamefont{Jausovec}}, \bibinfo{author}{\bibnamefont{O’Brien}}, \bibinfo{author}{\bibfnamefont{D.~E.} \bibnamefont{Read}}, \bibnamefont{and} \bibinfo{author}{\bibfnamefont{R.~P.} \bibnamefont{Cowburn}}, \bibinfo{journal}{Appl. Phys. Lett.} \textbf{\bibinfo{volume}{95}}, \bibinfo{pages}{152505} (\bibinfo{year}{2009}).

\bibitem[{\citenamefont{Thomas et~al.}(2006)\citenamefont{Thomas, Hayashi, Jiang, Moriya, Rettner, and Parkin}}]{CP1}
\bibinfo{author}{\bibfnamefont{L.}~\bibnamefont{Thomas}}, \bibinfo{author}{\bibfnamefont{M.}~\bibnamefont{Hayashi}}, \bibinfo{author}{\bibfnamefont{X.}~\bibnamefont{Jiang}}, \bibinfo{author}{\bibfnamefont{R.}~\bibnamefont{Moriya}}, \bibinfo{author}{\bibfnamefont{C.}~\bibnamefont{Rettner}}, \bibnamefont{and} \bibinfo{author}{\bibfnamefont{S.~S.~P.} \bibnamefont{Parkin}}, \bibinfo{journal}{Nature} \textbf{\bibinfo{volume}{443}}, \bibinfo{pages}{197} (\bibinfo{year}{2006}).

\bibitem[{\citenamefont{Nahrwold et~al.}(2009)\citenamefont{Nahrwold, Bocklage, Scholtyssek, Mat~suyama, Kr\"uger, Merkt, and Meier}}]{CP2}
\bibinfo{author}{\bibfnamefont{G.}~\bibnamefont{Nahrwold}}, \bibinfo{author}{\bibfnamefont{L.}~\bibnamefont{Bocklage}}, \bibinfo{author}{\bibfnamefont{J.~M.} \bibnamefont{Scholtyssek}}, \bibinfo{author}{\bibfnamefont{T.}~\bibnamefont{Mat~suyama}}, \bibinfo{author}{\bibfnamefont{B.}~\bibnamefont{Kr\"uger}}, \bibinfo{author}{\bibfnamefont{U.}~\bibnamefont{Merkt}}, \bibnamefont{and} \bibinfo{author}{\bibfnamefont{G.}~\bibnamefont{Meier}}, \bibinfo{journal}{J. Appl. Phys.} \textbf{\bibinfo{volume}{105}}, \bibinfo{pages}{07D511} (\bibinfo{year}{2009}).

\bibitem[{\citenamefont{Yershov et~al.}(2015)\citenamefont{Yershov, Kravchuk, Sheka, and Gaididei}}]{Yershov-Pinning}
\bibinfo{author}{\bibfnamefont{K.~V.} \bibnamefont{Yershov}}, \bibinfo{author}{\bibfnamefont{V.~P.} \bibnamefont{Kravchuk}}, \bibinfo{author}{\bibfnamefont{D.~D.} \bibnamefont{Sheka}}, \bibnamefont{and} \bibinfo{author}{\bibfnamefont{Y.}~\bibnamefont{Gaididei}}, \bibinfo{journal}{Phys. Rev. B} \textbf{\bibinfo{volume}{92}}, \bibinfo{pages}{104412} (\bibinfo{year}{2015}).

\bibitem[{\citenamefont{Bittencourt et~al.}(2022{\natexlab{b}})\citenamefont{Bittencourt, Castillo-Sep\'ulveda, Chubykalo-Fesenko, Moreno, Altbir, and Carvalho-Santos}}]{hilosElipticos}
\bibinfo{author}{\bibfnamefont{G.~H.~R.} \bibnamefont{Bittencourt}}, \bibinfo{author}{\bibfnamefont{S.}~\bibnamefont{Castillo-Sep\'ulveda}}, \bibinfo{author}{\bibfnamefont{O.}~\bibnamefont{Chubykalo-Fesenko}}, \bibinfo{author}{\bibfnamefont{R.}~\bibnamefont{Moreno}}, \bibinfo{author}{\bibfnamefont{D.}~\bibnamefont{Altbir}}, \bibnamefont{and} \bibinfo{author}{\bibfnamefont{V.~L.} \bibnamefont{Carvalho-Santos}}, \bibinfo{journal}{Phys. Rev. B} \textbf{\bibinfo{volume}{106}}, \bibinfo{pages}{174424} (\bibinfo{year}{2022}{\natexlab{b}}), \urlprefix\url{https://link.aps.org/doi/10.1103/PhysRevB.106.174424}.

\bibitem[{\citenamefont{Saitoh et~al.}(2004)\citenamefont{Saitoh, Miyajima, Yamaoka, and Tatara}}]{Saitoh2004}
\bibinfo{author}{\bibfnamefont{E.}~\bibnamefont{Saitoh}}, \bibinfo{author}{\bibfnamefont{H.}~\bibnamefont{Miyajima}}, \bibinfo{author}{\bibfnamefont{T.}~\bibnamefont{Yamaoka}}, \bibnamefont{and} \bibinfo{author}{\bibfnamefont{G.}~\bibnamefont{Tatara}}, \bibinfo{journal}{Nature} \textbf{\bibinfo{volume}{432}}, \bibinfo{pages}{203} (\bibinfo{year}{2004}), ISSN \bibinfo{issn}{1476-4687}, \urlprefix\url{https://doi.org/10.1038/nature03009}.

\bibitem[{\citenamefont{Sato et~al.}(2019)\citenamefont{Sato, Schultheiss, K\"orber, Puwenberg, M\"uhl, Awad, Arekapudi, Hellwig, Fassbender, and Schultheiss}}]{Sato-PRL}
\bibinfo{author}{\bibfnamefont{N.}~\bibnamefont{Sato}}, \bibinfo{author}{\bibfnamefont{K.}~\bibnamefont{Schultheiss}}, \bibinfo{author}{\bibfnamefont{L.}~\bibnamefont{K\"orber}}, \bibinfo{author}{\bibfnamefont{N.}~\bibnamefont{Puwenberg}}, \bibinfo{author}{\bibfnamefont{T.}~\bibnamefont{M\"uhl}}, \bibinfo{author}{\bibfnamefont{A.~A.} \bibnamefont{Awad}}, \bibinfo{author}{\bibfnamefont{S.~S. P.~K.} \bibnamefont{Arekapudi}}, \bibinfo{author}{\bibfnamefont{O.}~\bibnamefont{Hellwig}}, \bibinfo{author}{\bibfnamefont{J.}~\bibnamefont{Fassbender}}, \bibnamefont{and} \bibinfo{author}{\bibfnamefont{H.}~\bibnamefont{Schultheiss}}, \bibinfo{journal}{Phys. Rev. Lett.} \textbf{\bibinfo{volume}{123}}, \bibinfo{pages}{057204} (\bibinfo{year}{2019}).

\bibitem[{\citenamefont{Toro et~al.}(2020)\citenamefont{Toro, Alves, Carvalho-Santos, and de~Ara\'ujo}}]{Osc-JAP}
\bibinfo{author}{\bibfnamefont{O.~O.} \bibnamefont{Toro}}, \bibinfo{author}{\bibfnamefont{S.~G.} \bibnamefont{Alves}}, \bibinfo{author}{\bibfnamefont{V.~L.} \bibnamefont{Carvalho-Santos}}, \bibnamefont{and} \bibinfo{author}{\bibfnamefont{C.~I.~L.} \bibnamefont{de~Ara\'ujo}}, \bibinfo{journal}{J. Appl. Phys.} \textbf{\bibinfo{volume}{127}}, \bibinfo{pages}{183905} (\bibinfo{year}{2020}).

\bibitem[{\citenamefont{Sharma et~al.}(2015{\natexlab{b}})\citenamefont{Sharma, Muralidharan, and Tulapurkar}}]{Osc-Nat}
\bibinfo{author}{\bibfnamefont{S.}~\bibnamefont{Sharma}}, \bibinfo{author}{\bibfnamefont{B.}~\bibnamefont{Muralidharan}}, \bibnamefont{and} \bibinfo{author}{\bibfnamefont{A.}~\bibnamefont{Tulapurkar}}, \bibinfo{journal}{Sci. Rep.} \textbf{\bibinfo{volume}{5}}, \bibinfo{pages}{14647} (\bibinfo{year}{2015}{\natexlab{b}}).

\bibitem[{\citenamefont{Dhull et~al.}(2020)\citenamefont{Dhull, Nisar, and Kaushik}}]{Osc-Rev}
\bibinfo{author}{\bibfnamefont{S.}~\bibnamefont{Dhull}}, \bibinfo{author}{\bibfnamefont{A.}~\bibnamefont{Nisar}}, \bibnamefont{and} \bibinfo{author}{\bibfnamefont{B.~K.} \bibnamefont{Kaushik}}, in \emph{\bibinfo{booktitle}{Spintronics XIII}}, edited by \bibinfo{editor}{\bibfnamefont{H.-J.~M.} \bibnamefont{Drouhin}}, \bibinfo{editor}{\bibfnamefont{J.-E.} \bibnamefont{Wegrowe}}, \bibnamefont{and} \bibinfo{editor}{\bibfnamefont{M.}~\bibnamefont{Razeghi}}, \bibinfo{organization}{International Society for Optics and Photonics} (\bibinfo{publisher}{SPIE}, \bibinfo{year}{2020}), vol. \bibinfo{volume}{11470}, p. \bibinfo{pages}{114703Y}, \urlprefix\url{https://doi.org/10.1117/12.2568313}.

\bibitem[{\citenamefont{Ababei et~al.}(2021)\citenamefont{Ababei, Ellis, Vidamour, Devadasan, Allwood, Vasilaki, and Hayward}}]{Neuro1}
\bibinfo{author}{\bibfnamefont{R.~V.} \bibnamefont{Ababei}}, \bibinfo{author}{\bibfnamefont{M.~O.~A.} \bibnamefont{Ellis}}, \bibinfo{author}{\bibfnamefont{I.~T.} \bibnamefont{Vidamour}}, \bibinfo{author}{\bibfnamefont{D.~S.} \bibnamefont{Devadasan}}, \bibinfo{author}{\bibfnamefont{D.~A.} \bibnamefont{Allwood}}, \bibinfo{author}{\bibfnamefont{E.}~\bibnamefont{Vasilaki}}, \bibnamefont{and} \bibinfo{author}{\bibfnamefont{T.~J.} \bibnamefont{Hayward}}, \bibinfo{journal}{Sci. Rep.} \textbf{\bibinfo{volume}{11}}, \bibinfo{pages}{183905} (\bibinfo{year}{2021}).

\bibitem[{\citenamefont{Kumar et~al.}(2023)\citenamefont{Kumar, Chung, Chan, Jin, Lim, Parkin, Sbiaa, and Piramanayagam}}]{Neuro2}
\bibinfo{author}{\bibfnamefont{D.}~\bibnamefont{Kumar}}, \bibinfo{author}{\bibfnamefont{H.~J.} \bibnamefont{Chung}}, \bibinfo{author}{\bibfnamefont{J.}~\bibnamefont{Chan}}, \bibinfo{author}{\bibfnamefont{T.}~\bibnamefont{Jin}}, \bibinfo{author}{\bibfnamefont{S.~T.} \bibnamefont{Lim}}, \bibinfo{author}{\bibfnamefont{S.~S.~P.} \bibnamefont{Parkin}}, \bibinfo{author}{\bibfnamefont{R.}~\bibnamefont{Sbiaa}}, \bibnamefont{and} \bibinfo{author}{\bibfnamefont{S.~N.} \bibnamefont{Piramanayagam}}, \bibinfo{journal}{ACS Nano} \textbf{\bibinfo{volume}{17}}, \bibinfo{pages}{6261} (\bibinfo{year}{2023}).

\bibitem[{\citenamefont{Siddiqui et~al.}(2020)\citenamefont{Siddiqui, Dutta, Tang, Liu, Ross, and Baldo}}]{Neuro3}
\bibinfo{author}{\bibfnamefont{S.~A.} \bibnamefont{Siddiqui}}, \bibinfo{author}{\bibfnamefont{S.}~\bibnamefont{Dutta}}, \bibinfo{author}{\bibfnamefont{A.}~\bibnamefont{Tang}}, \bibinfo{author}{\bibfnamefont{L.}~\bibnamefont{Liu}}, \bibinfo{author}{\bibfnamefont{C.~A.} \bibnamefont{Ross}}, \bibnamefont{and} \bibinfo{author}{\bibfnamefont{M.~A.} \bibnamefont{Baldo}}, \bibinfo{journal}{Nano Lett.} \textbf{\bibinfo{volume}{20}}, \bibinfo{pages}{1033} (\bibinfo{year}{2020}).

\bibitem[{\citenamefont{Aharoni}()}]{Aharoni}
\bibinfo{author}{\bibfnamefont{A.}~\bibnamefont{Aharoni}}, \bibinfo{journal}{J. Appl. Phys.} \textbf{\bibinfo{volume}{83}}, \bibinfo{pages}{3432} (????).

\bibitem[{\citenamefont{Gaididei et~al.}(2017)\citenamefont{Gaididei, Goussev, Kravchuk, Pylypovskyi, Robbins, Sheka, Slastikov, and Vasylkevych}}]{Gaididei-JPA}
\bibinfo{author}{\bibfnamefont{Y.~B.} \bibnamefont{Gaididei}}, \bibinfo{author}{\bibfnamefont{A.}~\bibnamefont{Goussev}}, \bibinfo{author}{\bibfnamefont{V.~P.} \bibnamefont{Kravchuk}}, \bibinfo{author}{\bibfnamefont{O.~V.} \bibnamefont{Pylypovskyi}}, \bibinfo{author}{\bibfnamefont{J.~M.} \bibnamefont{Robbins}}, \bibinfo{author}{\bibfnamefont{D.~D.} \bibnamefont{Sheka}}, \bibinfo{author}{\bibfnamefont{V.}~\bibnamefont{Slastikov}}, \bibnamefont{and} \bibinfo{author}{\bibfnamefont{S.}~\bibnamefont{Vasylkevych}}, \bibinfo{journal}{J. Phys. A: Math. and Theor.} \textbf{\bibinfo{volume}{50}}, \bibinfo{pages}{385401} (\bibinfo{year}{2017}).

\bibitem[{\citenamefont{Yershov et~al.}(2018)\citenamefont{Yershov, Kravchuk, Sheka, Pylypovskyi, Makarov, and Gaididei}}]{Yershov-Auto}
\bibinfo{author}{\bibfnamefont{K.~V.} \bibnamefont{Yershov}}, \bibinfo{author}{\bibfnamefont{V.~P.} \bibnamefont{Kravchuk}}, \bibinfo{author}{\bibfnamefont{D.~D.} \bibnamefont{Sheka}}, \bibinfo{author}{\bibfnamefont{O.~V.} \bibnamefont{Pylypovskyi}}, \bibinfo{author}{\bibfnamefont{D.}~\bibnamefont{Makarov}}, \bibnamefont{and} \bibinfo{author}{\bibfnamefont{Y.}~\bibnamefont{Gaididei}}, \bibinfo{journal}{Phys. Rev. B} \textbf{\bibinfo{volume}{98}}, \bibinfo{pages}{060409(R)} (\bibinfo{year}{2018}).

\bibitem[{\citenamefont{Thiaville and Nakatani}(2006)}]{Thiavile}
\bibinfo{author}{\bibfnamefont{A.}~\bibnamefont{Thiaville}} \bibnamefont{and} \bibinfo{author}{\bibfnamefont{Y.}~\bibnamefont{Nakatani}}, in \emph{\bibinfo{booktitle}{Spin Dynamics in Confined Magnetic Structures III}}, edited by \bibinfo{editor}{\bibfnamefont{B.}~\bibnamefont{Hillebrands}} \bibnamefont{and} \bibinfo{editor}{\bibfnamefont{A.}~\bibnamefont{Thiaville}} (\bibinfo{publisher}{Springer}, \bibinfo{year}{2006}), p. \bibinfo{pages}{161}, ISBN \bibinfo{isbn}{978-3-540-20108-3}.

\bibitem[{\citenamefont{Slonczewski}(1972)}]{Slonczewski}
\bibinfo{author}{\bibfnamefont{J.~C.} \bibnamefont{Slonczewski}}, \bibinfo{journal}{AIP Conference Proceedings} \textbf{\bibinfo{volume}{5}}, \bibinfo{pages}{170} (\bibinfo{year}{1972}).

\bibitem[{\citenamefont{Kravchuk}(2014)}]{Kravchuk}
\bibinfo{author}{\bibfnamefont{V.}~\bibnamefont{Kravchuk}}, \bibinfo{journal}{J. Magn. Mag. Mat.} \textbf{\bibinfo{volume}{367}} (\bibinfo{year}{2014}).

\bibitem[{\citenamefont{Kr\"uger et~al.}(2007)\citenamefont{Kr\"uger, Pfannkuche, Bolte, Meier, and Merkt}}]{PhysRevB.75.054421}
\bibinfo{author}{\bibfnamefont{B.}~\bibnamefont{Kr\"uger}}, \bibinfo{author}{\bibfnamefont{D.}~\bibnamefont{Pfannkuche}}, \bibinfo{author}{\bibfnamefont{M.}~\bibnamefont{Bolte}}, \bibinfo{author}{\bibfnamefont{G.}~\bibnamefont{Meier}}, \bibnamefont{and} \bibinfo{author}{\bibfnamefont{U.}~\bibnamefont{Merkt}}, \bibinfo{journal}{Phys. Rev. B} \textbf{\bibinfo{volume}{75}}, \bibinfo{pages}{054421} (\bibinfo{year}{2007}), \urlprefix\url{https://link.aps.org/doi/10.1103/PhysRevB.75.054421}.

\end{thebibliography}

\end{document}